\documentclass[useAMS,usenatbib,usegraphicx]{mn2e}

\usepackage{url}
\usepackage[dvips]{color}

\newcommand{\ej}{\mathrm{ej}}
\newcommand{\csm}{\mathrm{csm}}
\newcommand{\kin}{\mathrm{kin}}
\newcommand{\s}{\mathrm{sh}}

\title[Mass-loss histories of SN IIn progenitors]
{
Mass-loss histories of Type IIn supernova progenitors
within decades before their explosion
}
\author[T. J. Moriya et al.]
{Takashi J. Moriya$^{1,2,3}$\thanks{moriyatk@astro.uni-bonn.de},
Keiichi Maeda$^{4,2}$,
Francesco Taddia$^5$,
Jesper Sollerman$^5$, 
\newauthor
Sergei I. Blinnikov$^{6,7,2}$, and
Elena I. Sorokina$^8$
 \\
$^{1}$
Argelander Institute for Astronomy, University of Bonn, Auf dem H\"ugel
71, D-53121 Bonn, Germany \\
$^{2}$
Kavli Institute for the Physics and Mathematics of the Universe (WPI),
Todai Institutes for Advanced Study,
University of Tokyo, \\ 5-1-5 Kashiwanoha, Kashiwa, Chiba 277-8583, Japan
\\
$^{3}$
Research Center for the Early Universe, Graduate School of Science,
University of Tokyo, 7-3-1 Hongo, Bunkyo, Tokyo, Japan\\
$^{4}$
Department of Astronomy, Kyoto University, Kitashirakawa-Oiwake, Sakyo, Kyoto 606-8502, Japan\\
$^{5}$
The Oskar Klein Centre, Department of Astronomy, Stockholm University,
AlbaNova, 10691 Stockholm, Sweden \\
$^{6}$
Institute for Theoretical and Experimental Physics, Bolshaya Cheremushkinskaya 25, 117218 Moscow, Russia\\
$^{7}$
Novosibirsk State University, Novosibirsk 630090, Russia\\
$^{8}$
Sternberg Astronomical Institute, M.V.Lomonosov Moscow State University,
Universitetski pr. 13, 119992 Moscow, Russia
}
\begin{document}

\date{Accepted 2014 January 17.  Received 2014 January 17; in original form 2013 December 18}

\pagerange{\pageref{firstpage}--11} \pubyear{2014}

\maketitle

\label{firstpage}

\begin{abstract}
We present results of a systematic study of the mass-loss properties
of Type IIn supernova progenitors within decades before their explosion.
We apply an analytic light curve model to
11 Type IIn supernova bolometric light curves 
to derive the circumstellar medium properties.
We reconstruct the mass-loss histories based on the estimated circumstellar medium
properties.
The estimated mass-loss rates are mostly higher than
$10^{-3}$ $M_\odot~\mathrm{yr^{-1}}$ and they are
 consistent with those obtained by other methods.
The mass-loss rates are often found to be constantly high
within decades before their explosion. This indicates that there exists
some mechanism to sustain the high mass-loss rates of Type IIn supernova progenitors
for at least decades before their explosion. Thus, the shorter eruptive mass loss
events observed in some Type IIn supernova progenitors 
are not always responsible for creating their dense circumstellar media.
In addition, we find that Type IIn supernova progenitors may tend to increase
their mass-loss rates as they approach to the time of their explosion.
We also show a detailed comparison between our analytic prediction and
numerical results.
\end{abstract}


\section{Introduction}\label{introduction}

Type IIn supernovae (SNe IIn) which were first named by
\citet{schlegel1990} are a subclass of SNe II.
They show narrow emission components in their spectra which are presumed to be
related to the existence of dense circumstellar media (CSM)
near the SN progenitors \citep[e.g.,][]{chugai1994,fransson2002}.
The existence of the dense CSM 
indicates that the SN IIn progenitors have high mass-loss rates 
shortly before the explosion. Indeed, some SNe IIn are related
to luminous blue variables (LBVs) which are at an evolutionary stage
of very massive stars \citep{humphreys1994}.
For instance, the progenitors of SNe IIn 2005gl, 2009ip, and 1961V are found to
be consistent with LBVs \citep[e.g.,][]{gal-yam2009,mauerhan2013,smith2011}.
However, LBVs have not been considered to be SN progenitors 
in the theoretical stellar evolution perspective \citep[e.g.,][]{langer2012},
although theoretical investigation of a possible LBV-like SN progenitor
is starting to appear \citep{groh2013}.
In addition, not all SNe IIn are related to very massive stars like
LBVs, but a large fraction of them may come from less massive stars
\citep[e.g.,][]{prieto2008,anderson2012}.

Estimating mass-loss histories of SN IIn progenitors is essential for
understanding their progenitors and mass-loss mechanisms.
Mass-loss rates of SN IIn progenitors have been estimated in many ways.
The line strength of H$\alpha$ in SNe IIn is an observational
property often used to estimate the CSM density and thus the mass-loss rate
(e.g., \citealt{taddia2013}, T13 hereafter; \citealt{stritzinger2012,kiewe2012}).
The dust emission observed in near- and mid-infrared has also been used 
(\citealt{maeda2013,fox2011,fox2013} and references therein).
X-ray observations are also widely used to estimate the CSM properties
\citep[e.g.,][]{dwarkadas2012,chandra2012,chandra2012b,katsuda2013}.
These observations commonly suggest that the mass-loss rates of SN IIn
progenitors are typically higher than $10^{-3}$ $M_\odot~\mathrm{yr^{-1}}$,
which is much higher than those estimated for other core-collapse SN progenitors
($\sim 10^{-5}$ $M_\odot~\mathrm{yr^{-1}}$ or less,
e.g., \citealt[][]{chevalier2006a,chevalier2006b}).

In a previous paper of ours (\citealt{moriya2013}, M13 hereafter),
we developed an analytic bolometric light-curve (LC) model for SNe IIn which can be used to
estimate the CSM properties.
We have applied our analytic model to the bolometric LCs reported
by \citet{stritzinger2012} (SNe 2005ip and 2006jd) and
\citet{zhang2012} (SN 2010jl) in M13.
In this paper, we additionally apply our bolometric LC model to those
reported by T13, \citet{fassia2000}, \citet{roming2012}, and \citet{fraser2013}.
In total, we estimate the mass-loss histories of 11 SN IIn progenitors and,
although the number is still small, we try to see if there are general
properties in the mass loss of SN IIn progenitors.

This paper is organized as follows.
In Section \ref{sec:model}, we shortly summarize our analytic bolometric
LC model
presented in M13. We apply the LC model to the bolometric LCs
in Section \ref{sec:history}. We summarize the results 
in Section \ref{sec:discussion} and see if
there exist general trends in the mass loss properties of SN IIn progenitors
shortly before their explosions. We present our conclusions in Section \ref{sec:conclusion}.
In Appendix \ref{sec:app}, we present a detailed comparison between our
analytic model and numerical results to demonstrate the reliability of our
analytic model.

\section{Bolometric Light Curve Model}\label{sec:model}
We briefly summarize the M13 analytic bolometric LC model for SNe whose major power
source is the interaction between SN ejecta and its CSM.
More detailed information on the model is presented in M13.

The analytic model assumes that the homologously-expanding
SN ejecta has two components in the density structure, $\rho_\ej\propto r^{-\delta}$ inside and
$\rho_\ej\propto r^{-n}$ outside, where $r$ is a radius. 
The parameter $n$ is known to be mostly determined
by the compactness of the progenitor from numerical simulations \citep[e.g.,][]{matzner1999}.
For example, explosions of red supergiants end up with $n\simeq 12$
while those of Wolf-Rayet stars lead to $n\simeq 10$.
The CSM is assumed to have a single power-law density structure [$\rho_\csm(r)=Dr^{-s}$].

We assume that the shocked SN ejecta and CSM form a thin shell because
of the efficient radiative cooling so that the shocked region can be expressed with
a single radius $r_\s(t)$, where $t$ is the time since the explosion.
Then, the evolution of the shell radius can be estimated through the
conservation of momentum, i.e.
\begin{equation}
 M_\s\frac{dv_\s}{dt}=4\pi r_\s^2 \left[\rho_\ej\left(v_\ej-v_\s\right)^2-\rho_\csm\left(v_\s-v_\csm\right)^2\right],
\label{momentum}
\end{equation}
where $M_\s$ is the total mass of the shocked SN ejecta and CSM,
$v_\s$ is the velocity of the shell, $v_\ej$ is the velocity of the SN ejecta
at $r_\s$, and $v_\csm$ is the CSM velocity.
The concrete form of Equation (\ref{momentum}) differs depending on 
the SN density structure entering the shell. 
At first, the outer SN ejecta with $\rho_\ej\propto r^{-n}$ enters the shell.
After a certain time $t_t$ when the region in the SN ejecta with
$\rho_\ej\propto r^{-n}$ is completely swept up, the density structure entering the
SN ejecta becomes $\rho_\ej\propto r^{-\delta}$.

Assuming that we get a solution for Equation (\ref{momentum}),
we are able to write down the kinetic energy $dE_\kin$ entering the
shell,
\begin{equation}
dE_\kin=4\pi r_\s^2 \frac{1}{2}\rho_\csm v_\s^2 dr_\s.
\end{equation}
If a fraction $\epsilon$ of the kinetic energy is transferred to
radiation energy, the SN bolometric luminosity $L$ can be expressed as
\begin{equation}
L=\epsilon \frac{dE_\kin}{dt}=2\pi \epsilon\rho_\csm r_\s^2 v_\s^3. \label{luminosity}
\end{equation}
In this paper, we assume $\epsilon=0.1$ if necessary as we assume in M13.

It turns out that the bolometric luminosity before $t=t_t$ has a simple power-law form
\begin{equation}
L=L_1t^\alpha,\label{Lbeforett}
\end{equation}
where $L_1$ is a constant and
\begin{equation}
\alpha = \frac{6s-15+2n-ns}{n-s}.
\end{equation}
If we can obtain $\alpha$ by fitting an observed SN bolometric LC before
$t_t$,
we can constrain the CSM density slope $s$ just from the bolometric LC by assuming $n$.
If there are spectral observations from which we can infer the shell
velocity evolution, we can estimate $D$ in
$\rho_\csm=Dr^{-s}$ and we can get information on the CSM density structure.
Even if there is no velocity information, we can still estimate $D$ by assuming the SN ejecta mass $M_\ej$ and energy $E_\ej$.

After $t=t_t$, there is no general analytic solution to Equation
(\ref{momentum}) and we do not have a simple expression for $L$.
However, we can solve an asymptotic form of Equation (\ref{momentum}) numerically
which is applicable at $t\gg t_t$. Then, we can use Equation (\ref{luminosity})
to estimate the bolometric luminosity.

Once we succeed in estimating the CSM density structure
$\rho_\csm=Dr^{-s}$, we can estimate the mass-loss rate evolution of
the SN progenitor by assuming a constant CSM velocity $v_\csm$.
This is simply because the CSM at $r$ is ejected at the time $t'=r/v_\csm$
before the explosion under this assumption. Then, 
the mass-loss history $\dot{M}(t')$ is
\begin{equation}
\dot{M}(t')=4\pi r^2\rho_\csm v_\csm=4\pi D v_\csm^{3-s} t'^{2-s},
\end{equation}
where $t'$ is the time before the explosion.

\section{Revealing Mass-Loss History}\label{sec:history}
We apply the bolometric LC model presented in M13 and summarized in the
previous section to the observed SN IIn bolometric LCs in this section.
The bolometric LCs of T13, \citet{fassia2000}, and \citet{fraser2013}
are constructed based on the photometric observations
covering from near-ultraviolet to near-infrared.
The bolometric LC of SN 2010jl constructed by \citet{zhang2012} is based only on
their optical photometric observations.
The bolometric LC of SN 2011ht is based on near-ultraviolet to optical
observations (\citealt{roming2012}, see also \citealt{pritchard2013}).

\subsection{SN 2005ip}
SN 2005ip has already been modeled in M13. The result of the LC fitting including the
statistical error is
\begin{equation}
L=(1.44\pm0.08)\times 10^{43}\left(\frac{t}{\mathrm{1~day}}\right)^{-0.536\pm0.013}\mathrm{erg~s^{-1}}.
\end{equation}
As shown in M13, $\alpha=-0.536\pm0.013$ corresponds to
$s=2.28\pm0.03$ ($n=10$) or $s=2.36\pm0.02$ ($n=12$).
Assuming $s=2.28$, the CSM density structure becomes
\begin{equation}
\rho_\csm(r)=8.4\times 10^{-16}\left(\frac{r}{10^{15}~\mathrm{cm}}\right)^{-2.28}
 \mathrm{g~cm^{-3}}.
\end{equation}
The corresponding mass-loss history is
\begin{equation}
\dot{M}(t')=2.3\times 10^{-3}
\left(\frac{v_\csm}{100~\mathrm{km~s^{-1}}}\right)^{0.72}
\left(\frac{t'}{1~\mathrm{year}}\right)^{-0.28} M_\odot~\mathrm{yr^{-1}},
\end{equation}
where $t'$ is the time before the explosion.
The mass-loss rate does not differ much whether we use $n=10$ or $n=12$.

\subsection{SN 2005kj}\label{secsn2005kj}
The bolometric LC of SN 2005kj declines so rapidly that
it is difficult to fit it by the M13 LC model.
However, we find that
the `shell-shocked diffusion' model (\citealt{smith2007,arnett1980}, but
see also \citealt{moriya2013c}) is consistent with the LC after the peak.
The possibility to apply the diffusion model to SNe IIn which cannot be
explained by the M13 model is further discussed in M13.
The diffusion model is applicable to the declining phase after the shock goes
through a dense CSM. The LC evolution is expressed as
\begin{equation}
L=L_0\exp\left[-\frac{t-t_p}{\tau_\mathrm{diff}}\left(1+\frac{t-t_p}{2\tau_\mathrm{exp}}\right)\right],
\label{shellshocked}
\end{equation}
where $t_p$ is the time of the maximum luminosity, $\tau_\mathrm{diff}$
is the characteristic diffusion timescale in the shocked dense CSM,
and $\tau_\mathrm{exp}$ is the expansion timescale of the shocked dense
CSM.

\begin{figure}
\begin{center}
 \includegraphics[width=\columnwidth]{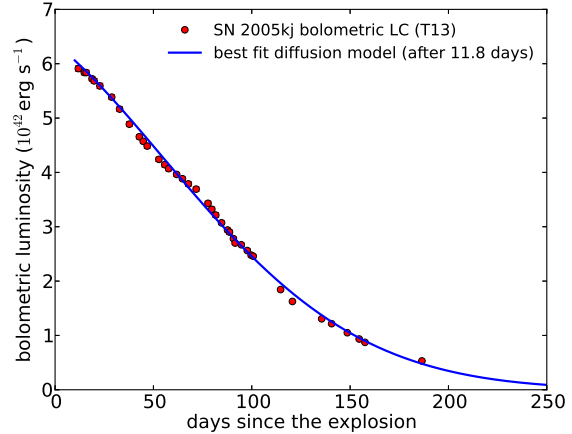}
  \caption{
Bolometric LC of SN 2005kj from T13 and a model fit to it based on the
 diffusion model. The diffusion model in the figure is obtained assuming
 that the LC peak is 10 days after the explosion.
}
\label{sn2005kj}
\end{center}
\end{figure}

Fig. \ref{sn2005kj} shows the result of the fitting to Equation
(\ref{shellshocked}) after the LC peak.
As the observed LC does not have a clear peak and the epoch of the
explosion is not well-determined,
we fit the LC assuming several possible $t_p$.
However, we found that the result of the fitting is not very sensitive
to the assumed $t_p$. In Fig. \ref{sn2005kj}, we show the LC obtained by
assuming that the first observed LC point is the LC peak
($t_p=11.8$ days since the explosion, T13). We then obtain
$\tau_\mathrm{diff}=175\pm9$ days, $\tau_\mathrm{exp}=57\pm5$ days, and
$L_0=(5.95\pm0.05)\times 10^{42}\ \mathrm{erg~s^{-1}}$.
Here, the errors are the statistical errors.
Assuming $t_p=5$ days, we obtain
$\tau_\mathrm{diff}=199\pm13$ days, $\tau_\mathrm{exp}=50\pm6$ days, and
$L_0=(6.17\pm0.06) \times 10^{42}\ \mathrm{erg~s^{-1}}$. 
In the most extreme case ($t_p=0$ days), we instead find
$\tau_\mathrm{diff}=221\pm17$ days, $\tau_\mathrm{exp}=45\pm5$ days, and
$L_0=(6.32\pm0.07) \times 10^{42}\ \mathrm{erg~s^{-1}}$.

We can roughly estimate the mass-loss rate of the progenitor from the
diffusion timescale. The diffusion timescale can be approximated
as
\begin{equation}
\tau_\mathrm{diff}\sim \frac{\kappa \overline{\rho} R^2}{c},
\end{equation}
where $\kappa$ is the opacity of the dense CSM,
$\overline{\rho}$ is the average density of the CSM,
$R$ is the radius of the CSM, and
$c$ is the speed of light. Then, the CSM mass $M_\csm$ is roughly
\begin{equation}
M_\csm = \frac{4}{3}\pi\overline{\rho}R^3\sim \frac{4\pi c
 \tau_\mathrm{diff}R}{3\kappa}.
\end{equation}
If $M_\csm$ is ejected in a time $\Delta t$, the mass-loss rate can be
approximated as $\dot{M}\sim M_\csm/\Delta t$. As $\Delta t=R/v_\csm$,
we get
\begin{equation}
\dot{M}\sim \frac{4\pi c\tau_\mathrm{diff}v_\csm}{3\kappa}.\label{diffrate}
\end{equation}
Using $\tau_\mathrm{diff}=175$ days, $v_\csm=100~\mathrm{km~s^{-1}}$
, and $\kappa=0.34\ \mathrm{cm^2~g^{-1}}$,
we get
\begin{equation}
\dot{M}\sim 0.9\ M_\odot~\mathrm{yr^{-1}}.
\end{equation}
This is a very rough estimate
but we can see that the mass-loss rate is high.

The fact that the LC can be fitted by the diffusion model indicates that the
dense part of the CSM is swept-up at early times, and thus the dense part of
the CSM is small in radius. As is discussed in M13, this is
naturally expected for the $s>3$ dense CSM.
However, the diffusion model just requires the existence of
the dense CSM near the progenitor.
Thus, we cannot constrain the possibility that
there exists a very dense $s<3$ CSM with a small radius.

\subsection{SN 2006aa}\label{secsn2006aa}
\begin{figure}
\begin{center}
 \includegraphics[width=\columnwidth]{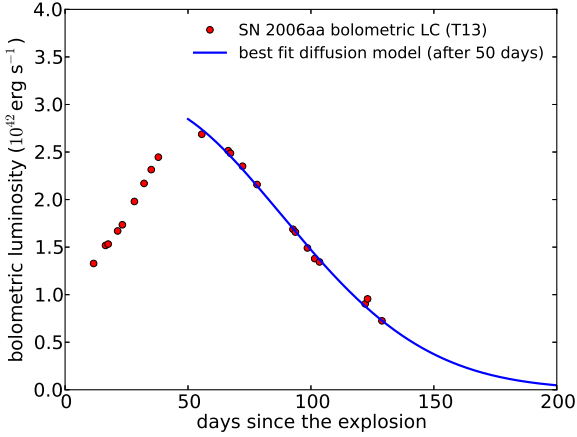}
  \caption{
Bolometric LC of SN 2006aa (T13) and its fit to the diffusion model.
The LC peak is assumed to be at 50 days since the explosion.
}
\label{sn2006aa}
\end{center}
\end{figure}

The bolometric LC of SN 2006aa could not be fitted by the M13 model as is the case for SN
2005kj. Again, the LC can be fitted by the diffusion model as shown in
Fig. \ref{sn2006aa}. We set the time of the peak luminosity as $t_p=50$ days
since the explosion and we get
$\tau_\mathrm{diff}=163\pm25$ days, $\tau_\mathrm{exp}=22\pm5$ days, and
$L_0=(2.85\pm0.05) \times 10^{42}\ \mathrm{erg~s^{-1}}$ with the
statistical errors. 
By using Equation (\ref{diffrate}), the mass-loss rate can be roughly
estimated as $\sim 0.8\ M_\odot~\mathrm{yr^{-1}}$.


\subsection{SN 2006bo}
\begin{figure}
\begin{center}
 \includegraphics[width=\columnwidth]{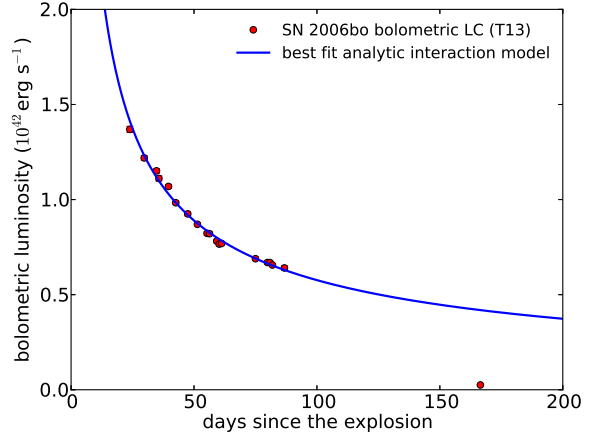}
  \caption{
Bolometric LC of SN 2006bo obtained in T13 and its fit to the
$L=L_1 t^\alpha$ formula.
}
\label{sn2006bo}
\end{center}
\end{figure}

The bolometric LC of SN 2006bo can be successfully fitted by the 
$L=L_1t^\alpha$ formula (Fig. \ref{sn2006bo}).
The explosion date is set as 20 days before the discovery but
it is not well-constrained (T13).
The result is
\begin{equation}
L=(1.03\pm0.06)\times 10^{43}\left(\frac{t}{\mathrm{1~day}}\right)^{-0.627\pm0.014}\mathrm{erg~s^{-1}}.
\end{equation}
The obtained $\alpha=-0.627\pm0.014$ corresponds to
$s=2.44\pm0.03$ ($n=10$) or $s=2.49\pm0.03$ ($n=12$).
The CSM density structure estimated for the $s=2.44$ case is
\begin{equation}
\rho_\csm(r)=2.5\times 10^{-15}\left(\frac{r}{10^{15}~\mathrm{cm}}\right)^{-2.44}\mathrm{g~cm^{-3}}.
\end{equation}
The Thomson optical depth above $10^{15}$ cm (the shell radius is mostly
above $10^{15}$ cm) for solar-metallicity CSM is 0.66 so our model
is self-consistent.
The mass-loss history estimated from
the CSM density structure is
\begin{equation}
\dot{M}(t')=9.1\times 10^{-3}
\left(\frac{v_\csm}{100~\mathrm{km~s^{-1}}}\right)^{0.56}
\left(\frac{t'}{1~\mathrm{year}}\right)^{-0.44} M_\odot~\mathrm{yr^{-1}},
\end{equation}
where $t'$ is the time before the explosion.
Note that we have ignored the bolometric luminosity data at around 170
days since the explosion when we fit the LC. The bolometric luminosity is
significantly smaller than the previous epochs.
As the bolometric LC is constructed by using near-infrared photometry as well,
this sudden luminosity decline is not necessarily from the dust formation.
We suspect that the shock has already gone out of the dense CSM at this epoch.
This indicates that the high mass-loss rate of the progenitor does not
last long enough for dense CSM to reach the corresponding radius (see
Section \ref{mlp}).

\subsection{SN 2006jd}
The CSM properties estimated from the bolometric LC of SN 2006jd were
presented in M13. The LC fitting with the $L=L_1 t^\alpha$ law
results in
\begin{equation}
L=(3.9\pm0.1)\times 10^{42}\left(\frac{t}{\mathrm{1~day}}\right)^{-0.0708\pm0.0064}\mathrm{erg~s^{-1}},
\end{equation}
with the statistical error.
The power $\alpha=-0.0708\pm0.0064$ indicates
$s=1.40\pm0.01$ ($n=10$) or $s=1.62\pm0.01$ ($n=12$).
The CSM density structure for the $s=1.40$ case is 
\begin{equation}
\rho_\csm(r)=2.6\times 10^{-16}\left(\frac{r}{10^{15}~\mathrm{cm}}\right)^{-1.40}~\mathrm{g~cm^{-3}}.
\end{equation}
The corresponding mass-loss rate is
\begin{equation}
\dot{M}(t')=2.6\times 10^{-4}
\left(\frac{v_\csm}{100~\mathrm{km~s^{-1}}}\right)^{1.6}
\left(\frac{t'}{1~\mathrm{year}}\right)^{0.6} M_\odot~\mathrm{yr^{-1}},
\end{equation}
where $t'$ is the time before the explosion.
The mass-loss rate decreases as the progenitor gets closer to the time
of explosion. The rate is higher than $10^{-3}$ $M_\odot~\mathrm{yr^{-1}}$
until about 9 years before the explosion.

\subsection{SN 2006qq}
\begin{figure}
\begin{center}
 \includegraphics[width=\columnwidth]{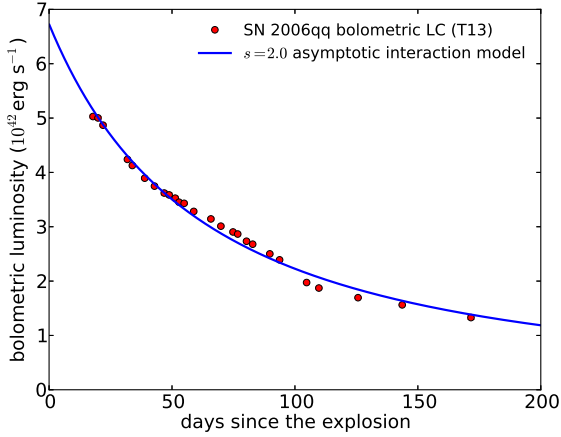}
  \caption{
Bolometric LC of SN 2006qq (T13) and its fit to the $s=2.0$ asymptotic
 LC model presented in M13. The $L=L_1 t^\alpha$ model fails because of
 a small $t_t$.
}
\label{sn2006qq}
\end{center}
\end{figure}

The bolometric LC of SN 2006qq cannot be fitted by the
$L=L_1t^\alpha$ model because of the small $t_t$. Thus, we use the
asymptotic formula to fit the LC.
The asymptotic formula does not generally have an analytic form.
We solve the equation numerically and see whether the fit is good or not
by eyes. We assume that the explosion date is 16 days before the discovery (T13).
We find that the $s=2.0$ asymptotic model provides a good fit (Fig. \ref{sn2006qq}).
Thus, we conclude that the mass loss of the progenitor is fully
consistent with being steady and we assign $s=2.0$ for SN 2006qq in the
following discussion.
The spectral observations of T13 indicate that the shock velocity is
almost constant with $10,000~\mathrm{km~s^{-1}}$, which is consistent
with the asymptotic model.
Including the velocity evolution, the CSM density structure is estimated as
\begin{equation}
\rho_\csm(r)=1.1\times 10^{-14}\left(\frac{r}{10^{15}~\mathrm{cm}}\right)^{-2.0}\mathrm{g~cm^{-3}}.
\end{equation}
The CSM optical depth becomes unity at around $3\times 10^{15}$ cm.
As $s=2.0$, the mass-loss rate of the progenitor is constant
\begin{equation}
\dot{M}(t')=2.1\times 10^{-2}
\left(\frac{v_\csm}{100~\mathrm{km~s^{-1}}}\right)
M_\odot~\mathrm{yr^{-1}}.
\end{equation}

\subsection{SN 2008fq}
\begin{figure}
\begin{center}
 \includegraphics[width=\columnwidth]{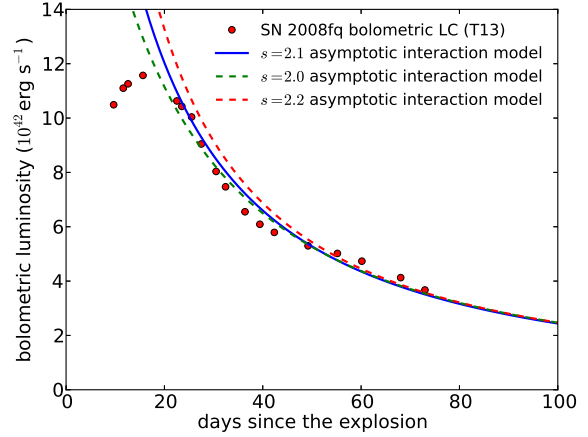}
  \caption{
Bolometric LC of SN 2008fq from T13 and asymptotic LC models for the LC.
The $L=L_1 t^\alpha$ model fails because of the small $t_t$.
Based on the LC models in the figure with several $s$, we conclude that $s=2.1$
provides the best fit to the LC after the peak.
}
\label{sn2008fq}
\end{center}
\end{figure}

Since the bolometric LC of SN 2008fq cannot be fitted by the $L=L_1t^{\alpha}$
formula self-consistently, 
we use the asymptotic one. We assume that the explosion date is
8 days before the discovery (T13).
If we assume $M_\ej=10~M_\odot$, the required CSM mass to fit the LC becomes
very large (about 50 $M_\odot$ within $10^{16}$ cm).
Hence, we assume $M_\ej=M_\odot$ instead for SN 2008fq.
Then, we find that
\begin{equation}
\rho_\csm(r)=3.8\times 10^{-14}\left(\frac{r}{10^{15}~\mathrm{cm}}\right)^{-2.1}\mathrm{g~cm^{-3}},
\end{equation}
with $E_\ej=1.3\times 10^{51}$ erg provides a better fit than the
$s=2.0$ or $s=2.2$ models (Fig. \ref{sn2008fq}).
Thus, we assign $s=2.1\pm0.05$ for SN 2008fq.
The Thomson optical depth above $10^{15}$ cm is 12.
The high optical depth is consistent with the existence of the early
long rise time when the photons emitted from the shell are presumed to
be scattered in the optically thick CSM.
The corresponding mass-loss history is
\begin{equation}
\dot{M}(t')=8.6\times 10^{-2}
\left(\frac{v_\csm}{100~\mathrm{km~s^{-1}}}\right)^{0.9}
\left(\frac{t'}{1~\mathrm{year}}\right)^{-0.1} M_\odot~\mathrm{yr^{-1}},
\end{equation}
where $t'$ is the time before the explosion.

\subsection{SN 2010jl}
The bolometric LC modeling of SN 2010jl is discussed in M13.
The bolometric LC is constructed by \citet{zhang2012} based on their
optical photometric observations.
The LC can be fitted by the $L=L_1t^\alpha$ law but $t_t$ becomes very
small and the $L=L_1t^\alpha$ model is not self-consistent.
We have applied the asymptotic model and we obtain the CSM density structure
\begin{equation}
\rho_\csm(r)=2.5\times
 10^{-14}\left(\frac{r}{10^{15}~\mathrm{cm}}\right)^{-2.2}
\mathrm{g~cm^{-3}}, \label{10jldensity}
\end{equation}
assuming $M_\ej=10~M_\odot$ (M13).
We again assign the statistical error of $0.05$ and we use $s=2.2\pm0.05$
for the SN 2010jl system in the next section.
The mass-loss rate derived from Equation (\ref{10jldensity}) is
\begin{equation}
\dot{M}(t')=6.2\times
 10^{-2}
\left(\frac{v_\csm}{100~\mathrm{km~s^{-1}}}\right)^{0.8}
\left(\frac{t'}{1~\mathrm{year}}\right)^{-0.2} M_\odot~\mathrm{yr^{-1}},
\end{equation}
where $t'$ is the time before the explosion.
The mass-loss rate recently reported by \citet{fransson2013} is within a
factor of a few $(0.11~M_\odot~\mathrm{yr^{-1}})$.
However, the mass-loss rate estimated by \citet{ofek2013b} is about one
order of magnitude higher and it is comparable to those of superluminous
SNe \citep[e.g.,][]{moriya2013b}.
This may be because \citet{ofek2013b} assume that the shock breakout occurred in
the dense CSM while we do not. The shock breakout requires very high
optical depth and thus, the large CSM mass.

\subsection{SN 2011ht}
\begin{figure}
\begin{center}
 \includegraphics[width=\columnwidth]{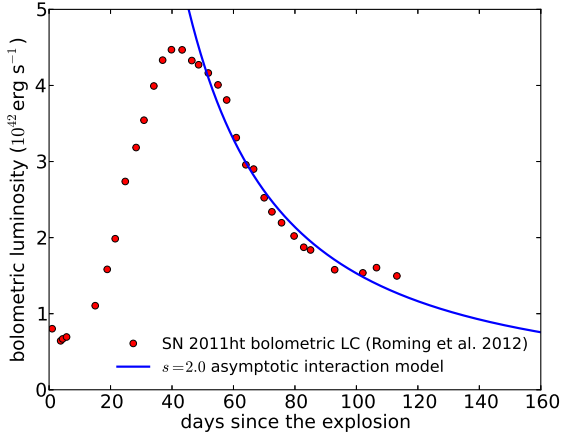}
  \caption{
Bolometric LC of SN 2011ht \citep{roming2012} and the $s=2.0$ asymptotic
 LC model for it.
}
\label{sn2011ht}
\end{center}
\end{figure}

The bolometric LC of SN 2011ht was constructed by \citet{roming2012}
based on their intensive near-ultraviolet and optical
observations and we use their bolometric LC for our modeling (see also \citealt{pritchard2013}).
A pre-SN burst was detected in one year before the explosion of SN 2011ht
\citep{fraser2013}.
There is a suggestion that SN 2011ht may not be a true core-collapse event
\citep{humphreys2012} but we assume it is.
The bolometric LC is shown in Fig. \ref{sn2011ht}.
In the first two observational epochs, the bolometric luminosity declines.
Thus, we assume that the first observed epoch is shortly after the
explosion and we set the explosion date one day before the first
observed epoch.

Fig. \ref{sn2011ht} shows the result of our bolometric LC fitting.
The $L=L_1 t^\alpha$ law does not work self-consistently and we use the
asymptotic form.
As the asymptotic $s=2.0$ model provides a good fit, we assign $s=2.0$ for SN
2011ht. The CSM density structure is constrained to
\begin{equation}
\rho_\csm (r)=5.0\times 10^{-15}\left(\frac{r}{10^{15}~\mathrm{cm}}\right)^{-2.0}~\mathrm{g~cm^{-3}}.
\end{equation}
The Thomson optical depth becomes unity at $5\times 10^{15}$ cm.
The corresponding mass-loss rate is
\begin{equation}
\dot{M}(t')=1.0\times 10^{-2}\left(\frac{v_\csm}{100~\mathrm{km~s^{-1}}}\right)\ M_\odot~\mathrm{yr^{-1}}.
\end{equation}
The estimated mass-loss rate is consistent with those estimated in the
previous studies, i.e.,
0.03 $M_\odot~\mathrm{yr^{-1}}$ \citep{mauerhan2013b} and
0.05 $M_\odot~\mathrm{yr^{-1}}$ \citep{humphreys2012}.

\subsection{SN 1998S}\label{secsn1998S}
The bolometric LC of SN 1998S is constructed by \citet{fassia2000} based
on their photometric observations in a wide spectral range.
They obtain two bolometric LCs for SN 1998S depending on the way they
fit the spectral energy distribution. We show their `spline' bolometric
LC instead of the `blackbody' one. The choice of the bolometric LC does
not affect our conclusion below.

\begin{figure}
\begin{center}
 \includegraphics[width=\columnwidth]{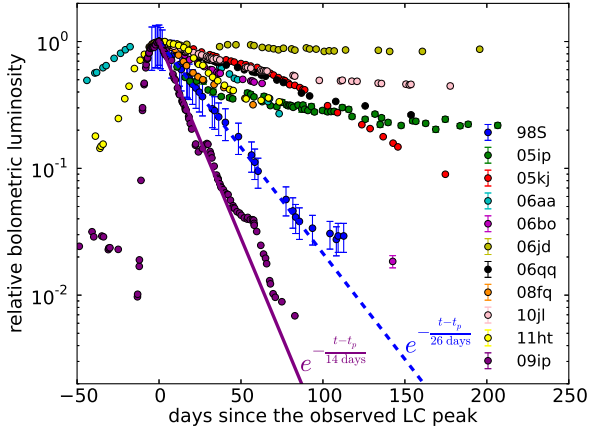}
  \caption{
Bolometric LCs of SN 1998S and SN 2009ip in 2012 compared with the other
 SN IIn LCs in this paper. The two LCs can be fitted by a single
 exponential function as indicated in the figure.
}
\label{komattachan}
\end{center}
\end{figure}

We find that the bolometric LC of SN 1998S declines much faster than
those we have modeled so far (Fig. \ref{komattachan}).
The LC in 100 days after the peak can be fitted by an exponential
function
\begin{equation}
 L=L_0 \exp\left(-\frac{t-t_p}{26\pm5~\mathrm{days}}\right).
\end{equation}
This LC form is expected in the diffusion model when
$\tau_\mathrm{diff}=26\pm 5$ days and
$t-t_p\ll 2 \tau_\mathrm{exp}$ (see Equation (\ref{shellshocked})).
This indicates that the expansion timescale of SN 1998S is very large.
This can be due to the efficient deceleration of the ejecta because of
the SN-CSM collision. The fast declining LC may also be related to
the asphericity of the dense CSM (Section \ref{asphericity}).

We can roughly estimate the mass-loss rate of the progenitor with
$\tau_\mathrm{diff}$ by Equation (\ref{diffrate}).
Assuming $v_\csm=100\ \mathrm{km~s^{-1}}$ and $\kappa=0.34\ \mathrm{cm^2~g^{-1}}$,
we obtain the rough mass-loss rate of $\dot{M}\sim0.01\ M_\odot~\mathrm{yr^{-1}}$.
The estimated mass-loss rate is rather high compared with those
estimated by the previous studies ($10^{-4}-10^{-3}\ M_\odot~\mathrm{yr^{-1}}$,
see \citealt{kiewe2012} for a summary).

\subsection{SN 2009ip}\label{secsn2009ip}
The major luminosity increase of SN 2009ip in 2012 was observed
intensively by many groups.
There is discussion about whether it is really a core-collapse event or not
\citep[e.g.,][]{smith2013b,fraser2013,pastorello2013,martin2013}
but here we
assume that the final brightening is due to a SN explosion.
We use the bolometric LC reported by \citet{fraser2013} for our modeling
(see also \citealt{margutti2013}).
As was the case for SN 1998S, the LC declines fast and it can be fitted by
an exponential function (Fig. \ref{komattachan}). The diffusion
timescale $\tau_\mathrm{diff}$ is $14\pm1$ days. The corresponding mass-loss
rate for the standard set of the parameters is
$\dot{M}\sim 9\times 10^{-3}$ $M_\odot~\mathrm{yr^{-1}}$. This mass-loss
rate is consistent with those estimated by \citet{fraser2013}
($10^{-2}-10^{-1}\ M_\odot~\mathrm{yr^{-1}}$) and
\citet{ofek2013c}
($10^{-3}-10^{-2}\ M_\odot~\mathrm{yr^{-1}}$).
\citet{baklanov2013} presents a LC model of SN
2009ip to demonstrate the dense shell method which is a newly-proposed
method to use SNe IIn as a primary standard candle
\citep[cf.][]{potashov2013}.
The CSM density slope is $s=3$ and
the average mass-loss rate is $10^{-2}$ $M_\odot~\mathrm{yr^{-1}}$
with $v_\csm=100\ \mathrm{km~s^{-1}}$, which is consistent with our result.

\section{Discussion}\label{sec:discussion}
We summarize the CSM properties and corresponding mass-loss histories of SN IIn
progenitors estimated in the previous section here.
We have applied our bolometric LC model to the observed ones until around
100-200 days since the explosion. 
The CSM shocked at these epochs are released from the progenitors
within about 30-60 years before their explosions and, the following
mass-loss histories we discuss correspond to those decades before the explosions.
This is because the typical CSM velocity of SN IIn progenitors
which is observed in very narrow P Cygni components of SNe IIn
is $\sim100~\mathrm{km~s^{-1}}$ (e.g., T13; \citealt{kiewe2012}),
while the typical shocked shell velocity is $\sim 10,000~\mathrm{km~s^{-1}}$.
As the SN shock propagates about 100 times faster, it should have taken
100 times longer for the CSM to reach the same radius.

\subsection{Overall mass-loss properties}\label{mlp}
\begin{figure}
\begin{center}
 \includegraphics[width=\columnwidth]{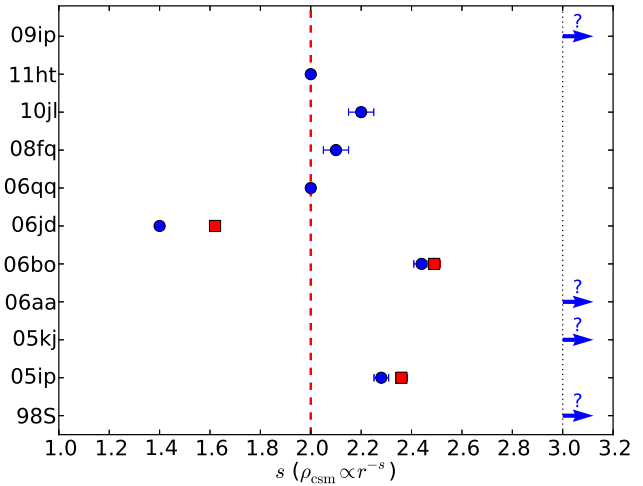}
  \caption{
Estimated CSM density slopes $s$ of the SN IIn progenitors.
When we need to assume $n$ to estimate $s$, we show the results of the
 cases of $n=10$ (blue circle) and $n=12$ (red square).
}
\label{sname}
\end{center}
\end{figure}

Fig. \ref{sname} summarizes the estimated CSM density slope $s$
($\rho_\csm\propto r^{-s}$). 
When we can fit the bolometric LCs by the $L=L_1t^\alpha$ law self-consistently,
we can estimate $s$ only from the bolometric LCs by assuming $n$.
In Fig. \ref{sname}, we plot two $s$,
one expected from $n=10$ (circle) and another from $n=12$ (square).
When we apply the asymptotic model,
we only plot one $s$ for each SN. 
There are four cases for which we apply the diffusion model. 
This may arise from $s>3$ CSM and we indicate these by arrows.

Looking at Fig. \ref{sname}, we can find that many $s$ gather around $2$.
This means that the mass loss of these SN IIn progenitors within decades before
the explosions is constantly large.
In other words, many SN IIn progenitors are likely to keep their
high mass-loss rates within the decades before their explosion.
In addition, there may exist a preference for $s$ to be larger than 2.
Assuming that the CSM velocity does not change much during the last
stage of the stellar evolution, the preference to $s>2$ means
that the mass-loss rates of SN IIn progenitors increase as the
progenitors get closer to the time of the explosion.
Note, however, that the systematic error is uncertain and can be
important. For instance, the uncertainty in
the estimated explosion dates is sometimes large.
Nonetheless,
the deviation from the steady mass loss in SN IIn CSM
has been suggested in previous studies as well.
For example, \citet{dwarkadas2012} collected 
SN X-ray LCs and found that SN IIn X-ray LCs are mostly not consistent
with the $s=2$ CSM. 

\begin{figure}
\begin{center}
 \includegraphics[width=\columnwidth]{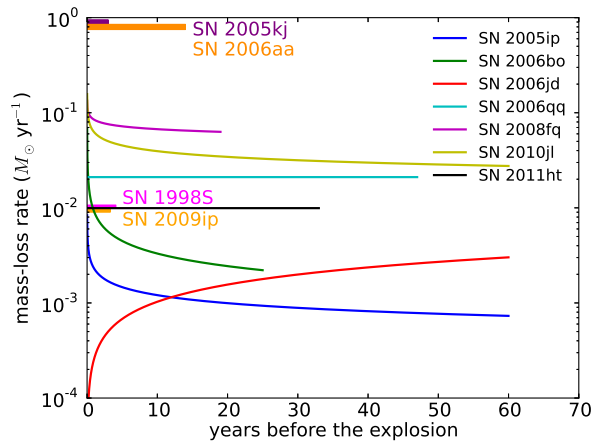}
  \caption{
Estimated mass-loss histories of the SN IIn progenitors.
The results of the $n=10$ models are shown when we need to assume $n$.
The shaded SNe (SNe 2005kj, 2006aa, 1998S, and 2009ip) are those
for which we roughly estimate the mass-loss rates based on the diffusion model.
$v_\csm=100$ $\mathrm{km~s^{-1}}$ is assumed in this figure.
}
\label{masslosshistory}
\end{center}
\end{figure}

Fig. \ref{masslosshistory} presents the history of the
mass-loss rates of SN IIn progenitors estimated in the previous section. 
We need to assume the SN ejecta properties in some cases and the uncertainty 
of the history is expected to be large.
The longest time we can trace depends on
 the time we used to fit the bolometric LCs.
For SNe IIn for which we apply the diffusion model, we indicate
the rough mass-loss rates estimated from the diffusion timescale
(Equation (\ref{diffrate})).
The longest time traced in these cases is set by assuming that the
entire dense CSM is swept up at the LC peak.
The mass-loss rates we obtain are consistent with
those obtained from other methods like H$\alpha$ luminosities which
also indicate that the mass-loss rates are typically higher than
$\sim 10^{-3}\ M_\odot~\mathrm{yr^{-1}}$ (e.g., T13; \citealt{kiewe2012}).

\subsection{Mass-loss mechanisms of SN IIn progenitors}
We have shown that the density slopes of dense CSM making SNe IIn
are often close to $s=2$. This indicates that the mass-loss rates
of SN IIn progenitors are constantly high within decades before their explosion.
In some SNe IIn, sudden luminosity increases of their progenitors
a few years to $\sim10$ days before their explosions
have been observed and are related to the formation of the dense CSM
\citep[e.g.,][]{pastorello2007,prieto2013,fraser2013b,ofek2013}.
SNe IIn for which we apply the diffusion model are mostly consistent with this
timescale (Fig. \ref{masslosshistory}) and these SN IIn progenitors
may make the dense CSM by the eruptive mass loss.
However, the progenitors of most SNe IIn we have modeled 
are found to have high mass-loss rates for decades. 
Thus, our results indicate that there exists some mechanism
for the progenitors to sustain their high mass-loss rates
at least for decades before their explosions.
The observed eruptive events on shorter timescales
do not explain all the dense CSM of SNe IIn.
Those eruptive events may make the dense CSM which are not smooth.
The existence of non-smooth dense CSM is indicated in some SN IIn
LCs which show short-time variability \citep[e.g., SN 2009ip,][]{margutti2013}.

We have also found that the mass-loss rates of SN IIn progenitors may preferentially
 get higher as they get closer to the time of the explosion.
However, mass loss which occurs at the surface of a star and the core collapse
which occurs at the center of the star are usually not physically connected to each other.
If the mass-loss rates of SN IIn progenitors truly tend to increase towards 
their time of the core collapse, this may indicate that the high
 mass-loss rates of SN IIn progenitors are somehow related to the core
 evolution of the progenitors.
There are several mechanisms to enhance the mass-loss rates which are
 triggered by the core evolution towards the core collapse and our result
 may support such mechanisms. 

For example, \citet{quataert2012,shiode2013} suggest that the g-mode
wave which is excited by the convective motion at the core 
 can convey energy to the surface.
 The conveyed energy can trigger the mass loss at the surface.
Since the convective motion can be more active as the nuclear burning
proceeds,
this mechanism may be able to explain the increasing mass-loss rates.
Although \cite{shiode2013} found that this mechanism may only work
 within about 10 years before the core collapse, it still remains a
 possible mechanism to enhance the mass loss towards the death.
Another example is the violent convective motion caused by the 
unstable nuclear burning \citep[e.g.,][]{smith2013}.
This mechanism may also be enhanced as the nuclear burning advances
since the advanced nuclear burning is more sensitive to temperature.

So far, we have emphasized the fact that the mass-loss rates of SN IIn
progenitors may tend to increase as they get closer to the time of the explosion.
However, there also exists an exception (SN 2006jd, see also \citealt{chandra2012}). In addition, the number of
SNe IIn we show here is small and we are still not at a stage of making
a strong statement from them. More SN IIn observations from which we can
estimate bolometric LCs and apply our LC model are required to get
a clearer view of the SN IIn mass loss.

\subsection{Effect of asphericity}\label{asphericity}
The bolometric LC model we applied to estimate the mass-loss rates
so far assumes the spherical symmetry.
However, the deviation from the spherical symmetry is reported in many
SNe IIn \citep[e.g.,][]{patat2011,trundle2009,leonard2000,levesque2014}.
In this section, we briefly discuss the effect of the asphericity
on the bolometric LCs and the mass-loss rates estimated in this paper.

The significant effect of the CSM asphericity on the bolometric LCs is in
the reduction of the dense CSM decelerating the SN ejecta.
The dense part of the CSM exists in all the directions in the spherically
symmetric case while the dense part only exists in some directions in
the aspherical case.
If the Thomson optical depth of the dense CSM is less than unity,
as is the case for the most SNe IIn we model here, the photons emitted
from the shock will be directly observed. Thus,
the bolometric luminosity is presumed to be roughly proportional
to the degree of the asymmetry for a given CSM density.
In other words, the bolometric luminosity
is expected to be reduced by the amount of the dense CSM decreased by
the asymmetry for a given CSM density. If the dense CSM with
$\rho_\csm = Dr^{-s}$ exists only at
the $\Omega$ direction out of $4\pi$, the average CSM density decreases to
$\langle\rho_\csm\rangle = \frac{\Omega}{4\pi} D r^{-s}$.
Here, we assume that the density of the sparse part of the CSM is significantly
smaller than that of the dense part.
The effect of the decrease in the average CSM density caused by the
asphericity on the luminosity is roughly included in the efficiency
$\epsilon$ in our
model. The efficiency will be decreased by $\frac{\Omega}{4\pi}$ by the
asymmetry because of the reduction of the average CSM density.
This means that the mass-loss rate of the $\Omega$ direction should be
increased by $\frac{4\pi}{\Omega}$ to get the same luminosity as the spherically
symmetric case so that the dense part of the CSM will be $\rho_\csm =\frac{4\pi}{\Omega} Dr^{-s}$.
However, even though the mass-loss rate should be increased in the
$\Omega$ direction to get the same luminosity,
the average mass-loss rate remains the same as the mass-loss rate
obtained by the spherically symmetric model because the average
density of the entire CSM becomes $\langle\rho_\csm\rangle=D r^{-s}$.
Thus, although the mass-loss rate of
a particular direction should be increased, the average mass-loss rate
is expected to remain roughly the same as the spherically symmetric case
to have the same luminosity in the aspherical case.

If the dense part of the CSM is optically thick, the effect of the
diffusion in the aspherical CSM is presumed to be significant and
the aspherical CSM can change the LCs more significantly, depending on
the viewing angle of the observers.
Some bolometric LCs shown in this paper are found to decline much
faster than those expected from the analytic model
(SNe 2005kj, 2006aa, 1998S, and 2009ip).
In the case of the optically thin CSM, the deviation from the spherical
symmetry is presumed to change the efficiency mainly without changing
the LC shape significantly. 
In addition, an interesting common property of the
fast-declining LCs is that their luminosities decline exponentially.
The exponential decay is not naturally expected from the asphericity of
the optically-thin CSM.
As we discussed earlier, the exponential decay is naturally expected from
the diffusion in the shocked optically-thick dense CSM.
Interestingly, the bolometric LCs of SNe 1998S and 2009ip, whose
LC declines are much faster than other SNe IIn, are suggested to
have a large asymmetry \citep{fassia2000,leonard2000,levesque2014}.
Since the diffusion process is significantly
affected by the asymmetry, these fast declines may be related to the
asphericity and the viewing angle of the observers.

We have discussed the possible effect of the deviation from the
spherical symmetry assumed in the analytic LC model qualitatively
in this section. However, the aspherical effect should be eventually
investigated quantitatively. We leave this as our future work.

\section{Conclusion}\label{sec:conclusion}
We have presented the results of our systematic study of the CSM around
SNe IIn.
To estimate the CSM properties,
we apply an analytic bolometric LC model for interacting SNe
formulated in M13 to 11 SN IIn bolometric LCs.
We have reconstructed the mass-loss histories of SN IIn progenitors
based on the estimated CSM properties.
As we typically use the bolometric LCs within 200 days since the
explosion, we are able to trace the mass-loss histories within about 60 years
before the explosion.

We find that mass-loss rates of many SNe IIn are constantly high
(above $\sim 10^{-3}$ $M_\odot~\mathrm{yr^{-1}}$) for more than a decade
before their explosion (Fig. \ref{masslosshistory}).
This suggests that the eruptive mass loss with shorter timescales
observed in several SN IIn
progenitors is not always a mechanism to make the dense CSM.
There should be a mass-loss mechanism which sustains the high mass-loss
rates at least for decades before the explosion.
In addition, we find that SN IIn progenitors
may tend to increase their mass-loss rates as they get closer to the
time of the explosion. If this is confirmed,
the currently unknown mass-loss mechanism of SN IIn progenitors may
be related to the core evolution of them. However,
the number of SNe IIn we modeled is still small and
we need more SN IIn observations from which we can construct bolometric LCs.

Revealing the progenitors of SNe IIn is important for the
 understanding of not only SNe but also stellar evolution.
SN IIn progenitors provide us with a clue to find missing keys in
the current stellar evolution theory. Some progenitor and mass-loss properties
are starting to be revealed as we show here in this paper.
However, we need more efforts to reach a better understanding of them.

\section*{Acknowledgements}
We would like to thank the referee for the comments which improved this paper.
T.J.M. is supported by the Japan Society for the Promotion of Science
Research Fellowship for Young Scientists (23\textperiodcentered 5929).
K.M. acknowledges the financial support by a Grant-in-Aid for
Scientific Research for Young Scientists (23740141).
This research is also supported by World Premier International Research Center Initiative (WPI Initiative), MEXT, Japan.
The Oskar Klein Centre is funded by the Swedish Research Council.
The work in Russia was supported by RF Government grant 11.G34.31.0047, grants for supporting Scientific Schools 5440.2012.2 and 3205.2012.2, and joint RFBR-JSPS grant 13--02--92119.

\appendix

\section{Comparison with Numerical Calculation}\label{sec:app}

In M13, some results of numerical LC calculations based on the initial
conditions obtained by the analytic model are presented.
Here, we show more detailed comparison between the analytic and numerical
models. We focus on the SN 2005ip model.
The parameters of the SN 2005ip progenitor system
estimated from the analytic model by assuming
$\delta=1$, $n=10$, $M_\ej=10~M_\odot$, and $\epsilon=0.1$ are 
$E_\ej=1.2\times 10^{52}$ erg and $\rho_\csm(r)=8.4\times
10^{-16}(r/10^{15}~\mathrm{cm})^{-2.28}$ $\mathrm{g~cm^{-3}}$.
We set the outer radius of the CSM at $5\times 10^{16}$ cm.
The numerical radiation hydrodynamics calculation is performed by
\texttt{STELLA}, which is a one-dimensional radiation hydrodynamics code
\citep[e.g.,][]{blinnikov2006,blinnikov1993}.
In \texttt{STELLA}, the conversion efficiency from kinetic energy to
radiation, which corresponds to $\epsilon$ in the analytic model,
is controlled by the smearing parameter
\citep{blinnikov1998,moriya2013b}.
We set the smearing parameter so that $\epsilon$ gets close to 0.1,
which is assumed in the analytic model.

\begin{figure}
\begin{center}
 \includegraphics[width=\columnwidth]{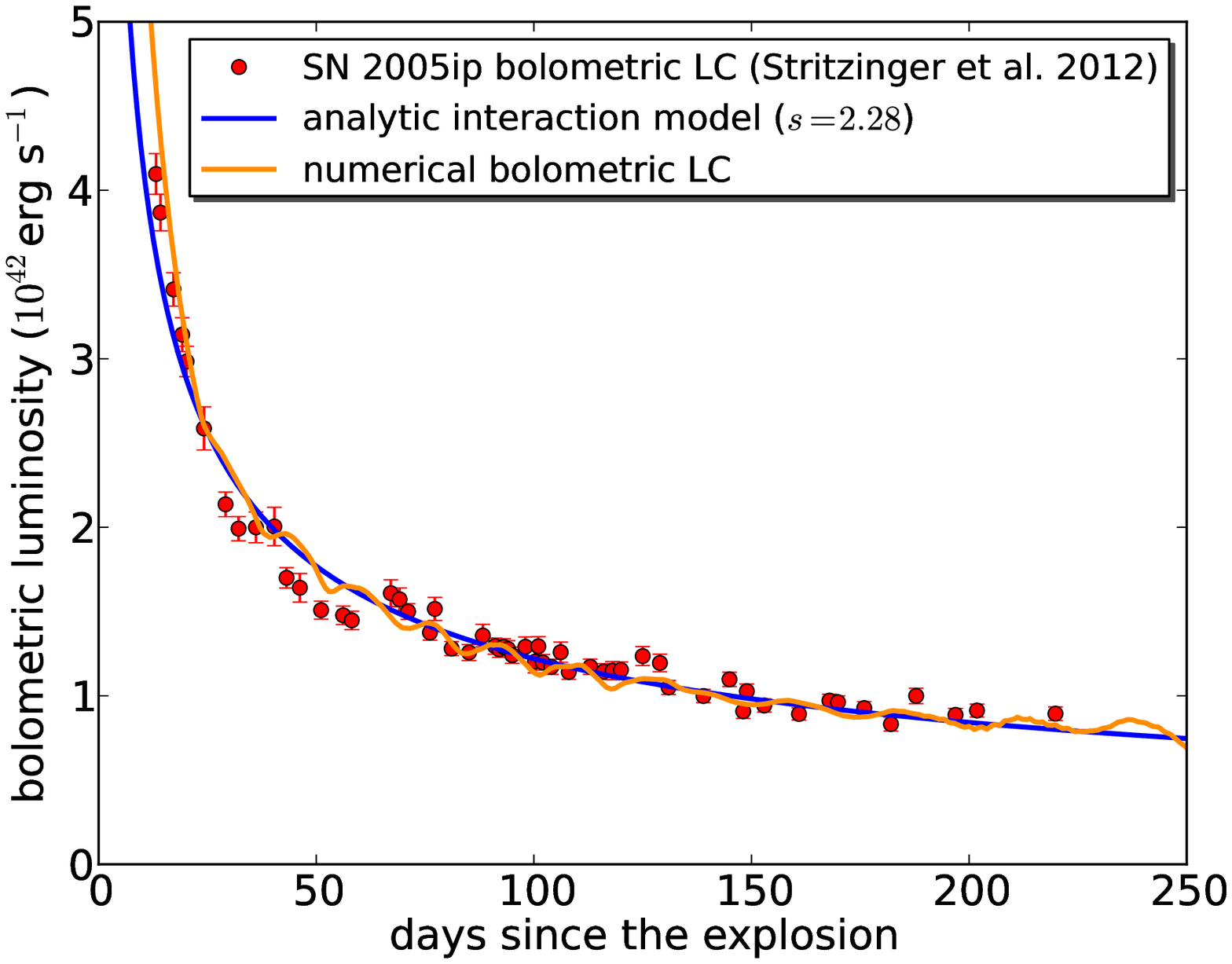}
  \caption{
Observed, analytic, and numerical LCs of SN 2005ip.
}
\label{sn2005ipLC}
\end{center}
\end{figure}

\begin{figure}
\begin{center}
 \includegraphics[width=\columnwidth]{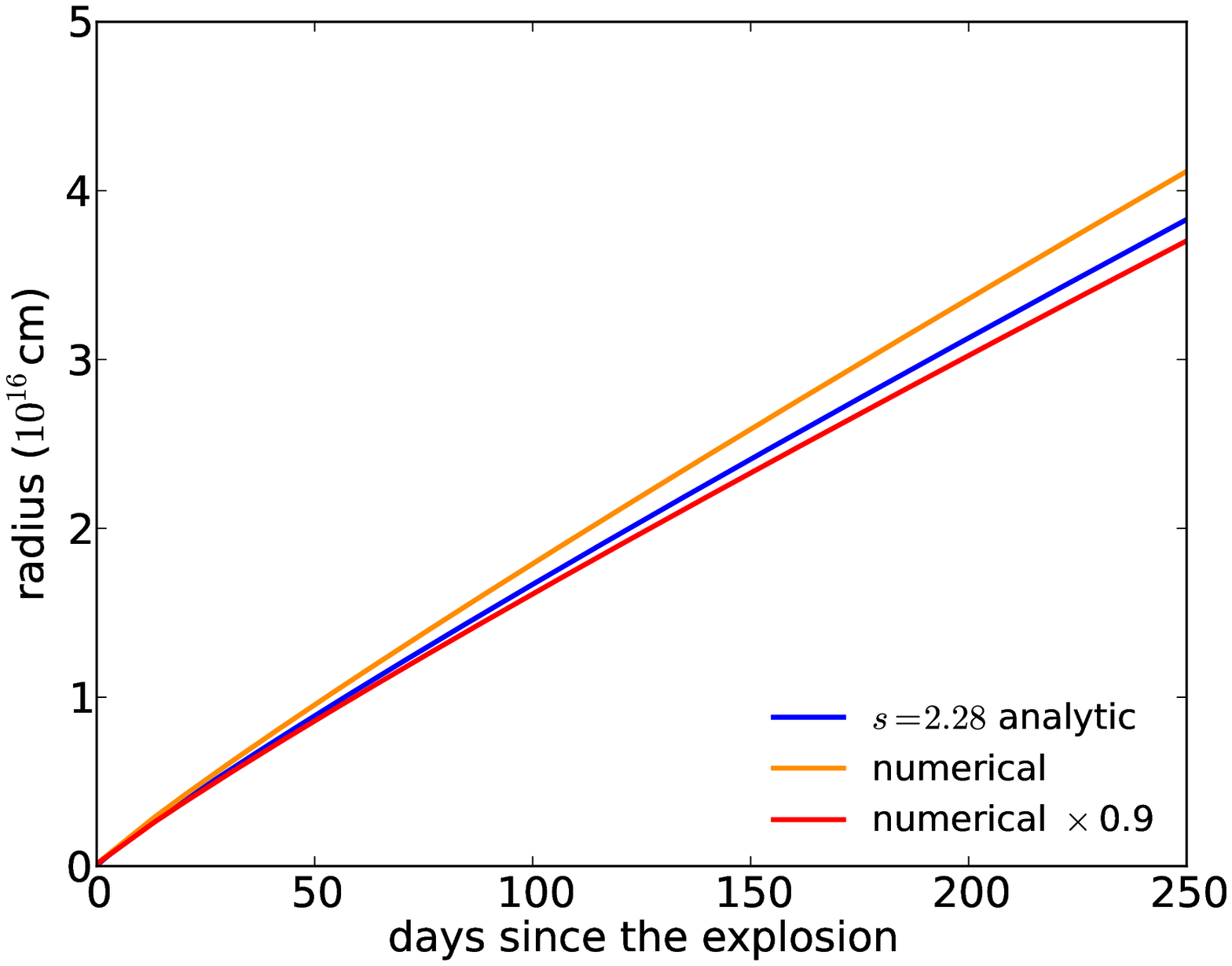}
  \caption{
Radial evolution obtained from the analytic model and that from
 numerical calculation. We also show a line which is 90\% of
 the numerical result.
}
\label{radius}
\end{center}
\end{figure}

\begin{figure}
\begin{center}
 \includegraphics[width=\columnwidth]{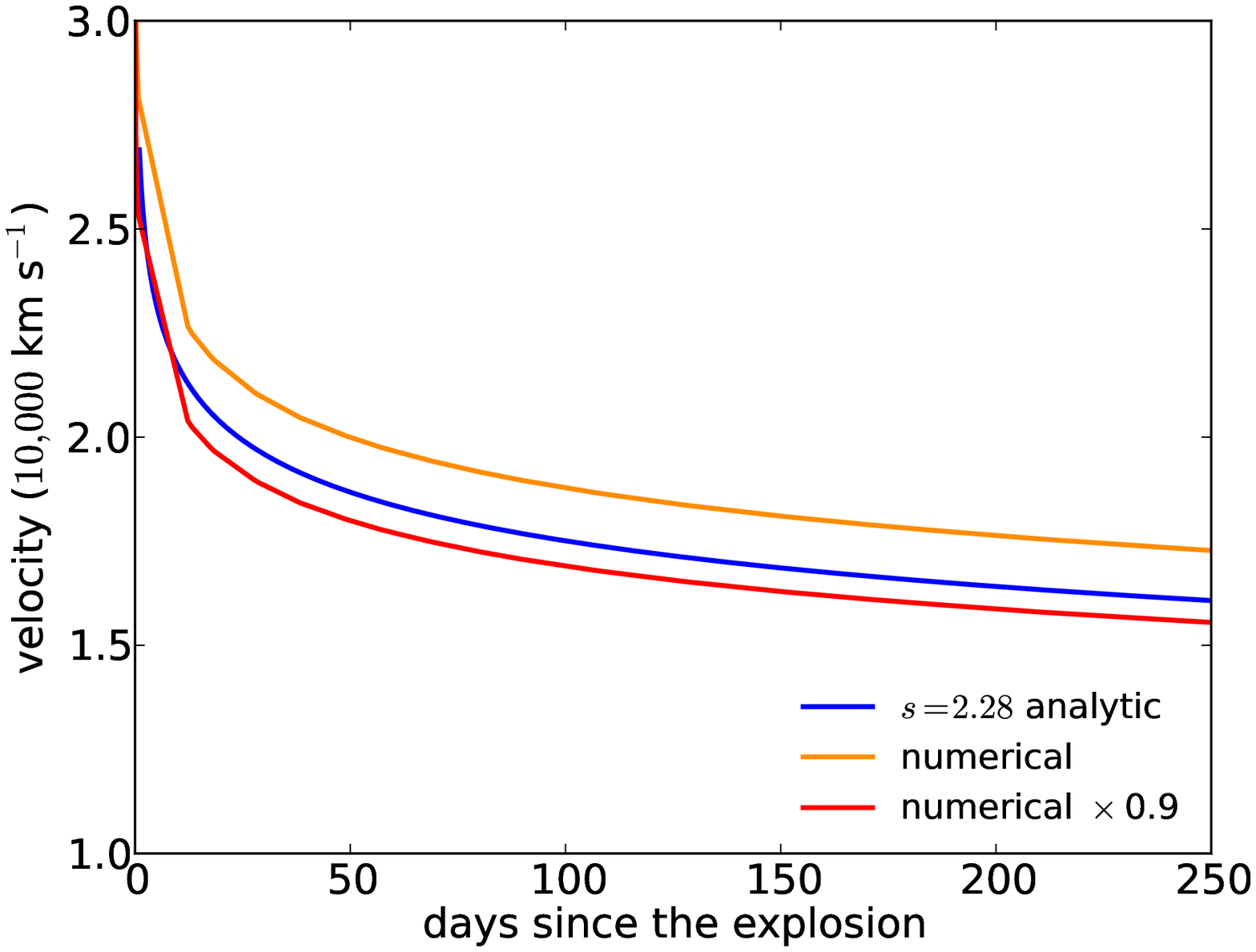}
  \caption{
The same as Fig. \ref{radius} but for the velocity.
}
\label{velocity}
\end{center}
\end{figure}

Fig. \ref{sn2005ipLC} shows the LCs obtained from the numerical calculation
and the analytic model. Overall, the two LCs match, although the
numerical LC is brighter until about 25 days since the explosion.
Fig.s \ref{radius} and \ref{velocity} compare the radii and velocities
of the numerical results to the analytic estimates.
Both the radius and velocity obtained from the
numerical result are a bit higher than the analytic
expectations. However, the difference is within 10 \%.

We finally estimate the conversion efficiency $\epsilon$ from the numerical
calculation.
The conversion efficiency is defined as 
\begin{equation}
\epsilon = L\left(\frac{dE_\kin}{dt}\right)^{-1}.
\end{equation}
$L$ is directly obtained by the numerical result.
To estimate $dE_\kin/dt$ from the numerical simulation, we assume that
the shell velocity $v_\s$ 
does not change much during a very small time
$\Delta t$. Then, by using the CSM mass $\Delta M$ swept up during $\Delta t$,
the total available kinetic energy $\Delta E_\kin$
during $\Delta t$ can be approximated as
\begin{eqnarray}
\Delta E_\kin &=& \frac{1}{2}\Delta M v_\s^2, \\
&\simeq& 2\pi D r_\s^{2-s} v_\s^3 \Delta t.
\end{eqnarray}
Then, $\epsilon$ can be approximated as
\begin{equation}
\epsilon\simeq L\left(\frac{\Delta E_\kin}{\Delta t}\right)^{-1}
=\frac{L}{2\pi D r_\s^{2-s} v_\s^3},\label{efficiencyeq}
\end{equation}
and we can estimate $\epsilon$ from $L$, $r_\s$, and $v_\s$, which are
available from the numerical calculation.

\begin{figure}
\begin{center}
 \includegraphics[width=\columnwidth]{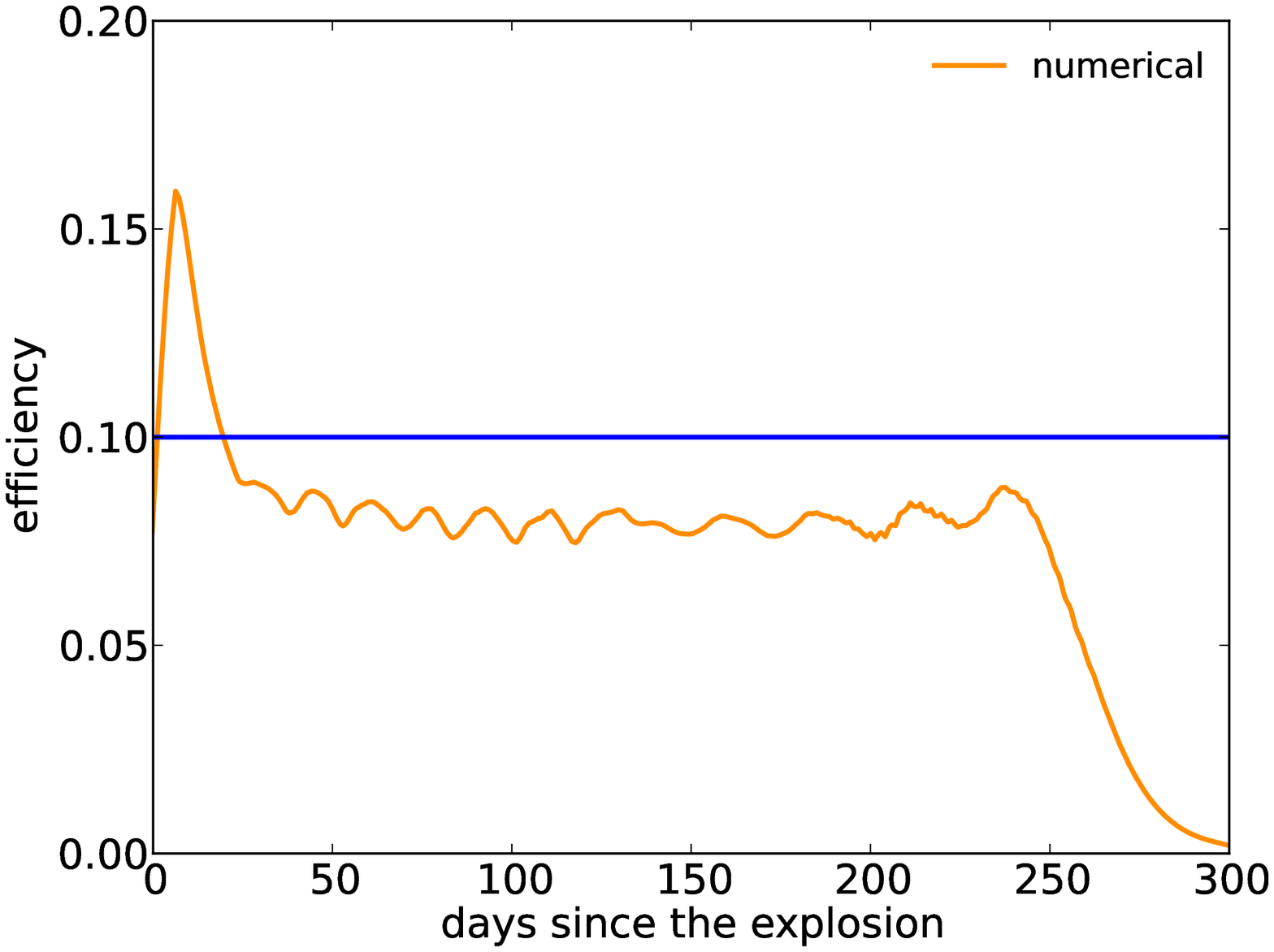}
  \caption{
Efficiency of the conversion from the available kinetic energy to
 radiation in the numerical model. The efficiency is estimated from
 Equation (\ref{efficiencyeq}).
}
\label{efficiency}
\end{center}
\end{figure}

Fig. \ref{efficiency} shows the efficiency obtained from the numerical
calculation. At early times, the efficiency gets large for a short
period of time but it becomes almost constant at around 0.1 later.
The assumption of the constant efficiency may not be valid in the early
times but it is a good approximation in most of time.
This means that the LC shape is mainly determined by the change in the
density, not by the change in the efficiency.

\label{lastpage}

\end{document}